\documentclass[conference, a4paper]{IEEEtran}

\usepackage{geometry}
\def\review{0}%
\def\arxiv{1}%
\usepackage[cmex10]{amsmath}
\usepackage{amssymb}
\usepackage{mathtools}
\usepackage{bm}
\usepackage{algorithmic}
\usepackage{graphicx}
\usepackage{xcolor}
\usepackage[acronym,toc]{glossaries}
\usepackage{siunitx}
\usepackage{tabularx}
\usepackage{booktabs}
\usepackage{pgfplots, tikz}
\usepackage{pgfplotstable}
\pgfplotsset{compat=1.18}
\usepgfplotslibrary{groupplots,fillbetween}
\usepackage{ifthen}
\usetikzlibrary{arrows.meta}
\usetikzlibrary{patterns}
\usepackage{layouts}
\usepackage{environ}
\usepackage{placeins}
\usepackage{subcaption}
\usepackage{stfloats}
\usepackage{textcase}
\usepackage[tablename=TABLE,font=small]{caption}
\usepackage{xcolor}
\usepackage{upgreek}
\usepackage{balance}
\usepackage{titlesec}
\usepackage{float}
\usepackage{times}
\usepackage[cmintegrals]{newtxmath}

\usepackage{cleveref}
\crefname{figure}{Fig.}{Figs.}
\Crefname{figure}{Fig.}{Figs.}
\crefname{subfigure}{Fig.}{Figs.}
\Crefname{subfigure}{Fig.}{Figs.}
\crefname{table}{Table}{Tables}
\Crefname{table}{Table}{Tables}
\crefname{equation}{Eq.}{Eqs.}
\Crefname{equation}{Eq.}{Eqs.}
\crefname{section}{Sec.}{Secs.}
\Crefname{section}{Sec.}{Secs.}
\usepackage[activate={true,nocompatibility},
    final,
    protrusion=false,
    tracking=true,
    kerning=true,
    spacing=true,
    stretch=25,
    shrink=25
    ]{microtype}

\makeatletter
\newsavebox{\measure@tikzpicture}
\NewEnviron{scaletikzpicturetowidth}[1]{%
    \def\tikz@width{#1}%
    \begin{lrbox}{\measure@tikzpicture}%
        \BODY
    \end{lrbox}%
    \pgfmathparse{#1/\wd\measure@tikzpicture}%
    \BODY
}
\makeatother

\usepackage{cite}
\ifthenelse{\equal{\review}{1}}{\usepackage{todonotes}}{\usepackage[disable]{todonotes}}

\newacronym{3gpp}{3GPP}{3rd generation partnership project}
\newacronym{5g}{5G}{fifth generation}
\newacronym{6g}{6G}{sixth generation}

\newacronym{admm}{ADMM}{alternating directions of multipliers method}
\newacronym{anm}{ANM}{atomic norm minimization}
\newacronym{adc}{ADC}{analog-to-digital converter}
\newacronym{aoa}{AoA}{angle of arrival}
\newacronym{aod}{AoD}{angle of departure}
\newacronym{awgn}{AWGN}{additive white Gaussian noise}
\newacronym{asic}{ASIC}{application specific integrated circuit}
\newacronym{arpack}{ARPACK}{Arnoldi package}
\newacronym{api}{API}{application programmable interface}
\newacronym{aut}{AUT}{antenna under test}
\newacronym[plural=AOI, firstplural=areas of interest (AOI)]{aoi}{AOI}{area of interest}
\newacronym{ai}{AI}{artifical intelligence}
\newacronym{aic}{AIC}{Akaike information criterion}

\newacronym{blue}{BLUE}{best linear unbiased estimator}
\newacronym{bp}{BP}{basis pursuit}
\newacronym{bpdn}{BPDN}{basis pursuit denoising}
\newacronym{brsrp}{BRSRP}{beam reference signal received power}
\newacronym{bs}{BS}{base station}

\newacronym{cnn}{CNN}{convolutional neural network}
\newacronym{crb}{CRB}{Cramér-Rao bound}
\newacronym{cs}{CS}{compressed sensing}
\newacronym{csi}{CSI}{channel state information}
\newacronym{cf}{CF}{crest factor}
\newacronym{cr}{CR}{compression ratio}
\newacronym{cmos}{CMOS}{complementary metal oxide semiconductor}
\newacronym{cots}{COTS}{commercial-off-the-shelf}
\newacronym{cp}{CP}{cyclic prefix}
\newacronym{cpu}{CPU}{central processing unit}
\newacronym{cfar}{CFAR}{constant false alarm rate}
\newacronym{ca-cfar}{CA-CFAR}{cell-averaging constant false alarm rate}
\newacronym{ca-cfar-short}{CA-CFAR}{cell-averaging CFAR}
\newacronym{osca-cfar}{OS-CA-CFAR}{ordered-statistic \glslink{ca-cfar}{cell-averaging CFAR}}
\newacronym{csca-cfar}{CS-CA-CFAR}{censored statistics \glslink{ca-cfar}{cell-averaging CFAR}}
\newacronym{so-cfar}{SO-CA-CFAR}{smallest-of \glslink{ca-cfar}{cell-averaging CFAR}}
\newacronym{cut}{CUT}{cell under test}
\newacronym{cvx}{CVX}{convex optimization toolbox}

\newacronym{dac}{DAC}{digital-to-analog converter}
\newacronym{doa}{DoA}{direction of arrival}
\newacronym{dod}{DoD}{direction of departure}
\newacronym[plural=DMC,firstplural=dense multipath components (DMC)]{dmc}{DMC}{dense multipath components}
\newacronym{das}{DAS}{delay-and-sum}
\newacronym{dft}{DFT}{discrete Fourier transform}
\newacronym{idft}{IDFT}{inverse discrete Fourier transform}
\newacronym{dct}{DCT}{discrete cosine transform}
\newacronym{dlearn}{DL}{deep learning}
\newacronym{dl}{DL}{downlink}
\newacronym{dsp}{DSP}{digital signal processor}
\newacronym{dnn}{DNN}{deep neural network}
\newacronym{dpss}{DPSS}{discrete prolate spheroidal sequences}

\newacronym{eadf}{EADF}{effective aperture distribution function}
\newacronym{etadf}{ETADF}{effective time-aperture distribution function}
\newacronym{esprit}{ESPRIT}{estimation of signal parameters via rotational invariance techniques}
\newacronym{ett}{ETT}{eigenvalue threshold test}
\newacronym{edc}{EDC}{efficient detection criterion}
\newacronym{eft}{EFT}{exponential fitting test}

\newacronym{fa}{FA}{false alarm}
\newacronym{fc}{FC}{fully-connected}
\newacronym{fht}{FHT}{fast Hadamard transform}
\newacronym[longplural={fast Fourier transforms}]{fft}{FFT}{fast Fourier transform}
\newacronym{fmcw}{FMCW}{frequency-modulated continuous-wave}
\newacronym{fpga}{FPGA}{field programmable gate array}
\newacronym{fri}{FRI}{finite rate of innovation}
\newacronym{fim}{FIM}{Fisher information matrix}
\newacronym{fmc}{FMC}{full matrix capture}
\newacronym{fn}{FN}{false negative}
\newacronym{fista}{FISTA}{fast iterative shrinkage-thresholding algorithm}
\newacronym{fp}{FP}{false positive}
\newacronym{fr2}{FR2}{frequency range 2}
\newacronym{frvm}{FRVM}{fast relevance vector machine}

\newacronym{gan}{GAN}{generative adversarial network}
\newacronym{gfcs}{grid-free CS}{grid-free compressive sensing}
\newacronym{gnb}{gNB}{gNodeB}
\newacronym{gm}{GM}{Gaussian mixture}
\newacronym{go}{GO}{greatest-of}
\newacronym{gpu}{GPU}{graphical processing unit}
\newacronym{gtd}{GTD}{geometrical theory of diffraction}

\newacronym{hrpe}{HRPE}{high resolution parameter estimator}
\newacronym{hfss}{Ansys HFSS}{Ansys high frequency electromagnetic simulation software}

\newacronym[longplural={inverse fast Fourier transforms}]{ifft}{IFFT}{inverse fast Fourier transform}
\newacronym{ir}{IR}{impulse response}
\newacronym{iid}{iid}{independent and identically distributed}
\newacronym{iir}{IIR}{infinite impulse response}
\newacronym{irf}{IRF}{impulse response function}
\newacronym{isac}{ISAC}{integrated sensing and communications}
\newacronym{ista}{ISTA}{iterative shrinkage-thresholding algorithm}

\newacronym{kf}{KF}{Kalman filter}
\newacronym{kld}{KLD}{Kullback-Leibler divergence}

\newacronym{ls}{LS}{least squares}
\newacronym{lasso}{LASSO}{least absolute shrinkage and selection operator}
\newacronym{lse}{LSE}{line spectral estimation}
\newacronym{lfsr}{LFSR}{linear feedback shift register}
\newacronym{lo}{LO}{local oscillator}
\newacronym{los}{LOS}{line-of-sight}
\newacronym{lti}{LTI}{linear time-invariant}
\newacronym{lam}{LAM}{large area monitoring}

\newacronym{mimo}{MIMO}{multiple input multiple output}
\newacronym{mmv}{MMV}{multiple measurement vectors}
\newacronym{ml}{ML}{maximum likelihood}
\newacronym{mmse}{MMSE}{misspecified mean squared error}
\newacronym{mmwave}{mmWave}{millimeter wave}
\newacronym{mse}{MSE}{mean squared error}
\newacronym{bce}{BCE}{binary crossentropy}
\newacronym{mlbs}{MLBS}{maximum length binary sequence}
\newacronym{mo}{MO}{modulation order}
\newacronym{mpc}{MPC}{multipath component}
\newacronym{msm}{MSM}{M-sequence method}
\newacronym{mwc}{MWC}{modulated wideband converter}
\newacronym{mpm}{MPM}{matrix pencil method}
\newacronym{mpu}{MPU}{microprocessor unit}
\newacronym{mri}{MRI}{magnetic resonance imaging}
\newacronym{music}{MUSIC}{multiple signal classification}
\newacronym{mkl}{MKL}{math kernel library}
\newacronym{mcrb}{MCRB}{misspecified Cramér-Rao bound}
\newacronym{mmle}{MMLE}{misspecified maximum-likelihood estimator}

\newacronym{ndt}{NDT}{nondestructive testing}
\newacronym{nde}{NDE}{nondestructive evaluation}
\newacronym{nlos}{NLOS}{non-line-of-sight}
\newacronym{nneigh}{NN}{nearest neighbor}
\newacronym{nist}{NIST}{National Institute of Standards and Technology}

\newacronym{ofdm}{OFDM}{orthogonal frequency-division multiplexing}
\newacronym{omp}{OMP}{orthogonal matching pursuit}
\newacronym{oop}{OOP}{object oriented programming}

\newacronym{pa}{PA}{power amplifier}
\newacronym{pdf}{pdf}{probability density function}
\newacronym{pdp}{PDP}{power delay profile}
\newacronym{phd}{PHD}{probability hypothesis density}
\newacronym{pn}{PN}{pseudo-noise}
\newacronym{poc}{PoC}{proof of concept}
\newacronym{prs}{PRS}{positioning reference signal}
\newacronym{re}{RE}{resource element}
\newacronym{pwc}{PWC}{plane wave compounding}
\newacronym{pwi}{PWI}{plane wave imaging}
\newacronym{pura}{PURA}{patch uniform rectangular array}
\newacronym{pymax}{PyMAX}{Python maximization approach}
\newacronym{psf}{PSF}{point spread function}
\newacronym{pts}{PTS}{pseudo-true solution}

\newacronym{qpsk}{QPSK}{Quadrature phase-shift keying}

\newacronym{ram}{RAM}{random access memory}
\newacronym{rc}{RC}{raised cosine}
\newacronym{rcs}{RCS}{radar cross section}
\newacronym{rd}{RD}{random demodulator}
\newacronym{relu}{ReLU}{rectified linear unit}
\newacronym{rem}{REM}{reconstruction error metric}
\newacronym{resnet}{ResNet}{residual neural network}
\newacronym{ric}{RIC}{restricted isometry constant}
\newacronym{rimax}{RIMAX}{Richter maximization approach}
\newacronym{rf}{RF}{radio frequency}
\newacronym{rip}{RIP}{restricted isometry property}
\newacronym{rms}{RMS}{root mean squared}
\newacronym{rmse}{RMSE}{root mean squared error}
\newacronym{roc}{ROC}{receiver operating characteristic}
\newacronym{roi}{ROI}{region of interest}
\newacronym{rsrp}{RSRP}{reference signal received power}
\newacronym{rs}{RS}{reference signal}
\newacronym{rvm}{RVM}{relevance vector machine}
\newacronym{ru}{RU}{radio unit}
\newacronym{rx}{RX}{receiver}

\newacronym{scf}{SCF}{spatial correlation function}
\newacronym{samurai}{SAMURAI}{synthetic aperture measurements of uncertainty in angle of incidence}
\newacronym[plural=SC,firstplural=specular components (SC)]{sc}{SC}{specular components}
\newacronym{sdp}{SDP}{semi-definite program}
\newacronym{sdr}{SDR}{signal to diffuse ratio}
\newacronym{simd}{SIMD}{single instruction multiple data}
\newacronym{svd}{SVD}{singular value decomposition}
\newacronym{svm}{SVM}{support vector machine}
\newacronym{soe}{SOE}{sparsity order estimation}
\newacronym{sgd}{SGD}{stochastic gradient descent}
\newacronym{stuca}{StUCA}{stacked uniform circular array}
\newacronym{slam}{SLAM}{simultaneous localization and mapping}
\newacronym{spucpa}{SPUCPA}{stacked polarimetric uniform circular patch array}
\newacronym{suca}{SUCA}{stacked uniform circular array}
\newacronym{saft}{SAFT}{synthetic aperture focusing technique}
\newacronym{sota}{SOTA}{state of the art}
\newacronym{ssd}{SSD}{solid state device}
\newacronym{ssr}{SSR}{sparse signal recovery}
\newacronym{sa}{SA}{synthetic aperture}
\newacronym{spw}{SPW}{single plane wave}
\newacronym{shm}{SHM}{structural health monitoring}
\newacronym{snr}{SNR}{signal-to-noise ratio}
\newacronym{srrc}{SRRC}{square-root raised cosine}
\newacronym{stela}{STELA}{soft-thresholding with exact line search algorithm}
\newacronym{siso}{SISO}{single input single output}
\newacronym{simo}{SIMO}{single input multiple output}
\newacronym{swe}{SWE}{spherical wave expansion}
\newacronym{sme}{SME}{spherical mode expansion}
\newacronym{sage}{SAGE}{space-alternating generalized expectation-maximization}

\newacronym{tdd}{TDD}{time division duplex}
\newacronym{th}{T\&H}{Track and Hold}
\newacronym{tf}{TF}{transfer function}
\newacronym{tof}{ToF}{time-of-flight}
\newacronym{toa}{ToA}{time of arrival}
\newacronym{toc}{TOC}{tracker operating characteristic}
\newacronym{tn}{TN}{true negative}
\newacronym{tp}{TP}{true positive}
\newacronym{twista}{TWISTA}{two-step iterative shrinkage-thresholding algorithm}
\newacronym{tx}{TX}{transmitter}

\newacronym{uca}{UCA}{uniform circular array}
\newacronym{ue}{UE}{user equipment}
\newacronym{ura}{URA}{uniform rectangular array}
\newacronym{ul}{UL}{uplink}
\newacronym{ula}{ULA}{uniform linear array}
\newacronym{upa}{UPA}{uniform planar array}
\newacronym{uwb}{UWB}{ultra-wideband}
\newacronym{usndt}{US-NDT}{ultrasonic non-destructive testing}

\newacronym{vna}{VNA}{vector network analyser}
\newacronym{vsh}{VSH}{vector spherical harmonics}
\makeglossaries
\def\BibTeX{{\rm B\kern-.05em{\sc i\kern-.025em b}\kern-.08em
T\kern-.1667em\lower.7ex\hbox{E}\kern-.125emX}}

\newcommand{\N}{\mathbb{N}}

\newcommand{\E}{\bm{\mathrm{E}}}

\newcommand\blfootnote[1]{%
  \begingroup
  \renewcommand\thefootnote{}\footnote{#1}%
  \addtocounter{footnote}{-1}%
  \endgroup
}

\definecolor{QC1}{HTML}{0077BB}
\definecolor{QC2}{HTML}{33BBEE}
\definecolor{QC3}{HTML}{009988}
\definecolor{QC4}{HTML}{EE7733}
\definecolor{QC5}{HTML}{CC3311}
\definecolor{QC6}{HTML}{EE3377}
\definecolor{QC7}{HTML}{BBBBBB}
\definecolor{QCW}{HTML}{000000}

\definecolor{palPath}{RGB}{217,84,26}%
\definecolor{palPathDark}{RGB}{130,50,16}%
\definecolor{palNoise}{RGB}{0,115,188}%
\definecolor{palNoiseDark}{RGB}{0,69,113}%

\ifx\qty\undefined
    \let\oldnum\num
    \renewcommand{\num}[2][ignored]{\oldnum{#2}}

    \ifx\qty\undefined
        \newcommand{\qty}[3][ignored]{\SI{#2}{#3}}
    \fi

    \ifx\qtyrange\undefined
        \newcommand{\qtyrange}[4][ignored]{\SIrange{#2}{#3}{#4}}
    \fi
\fi

\usepackage{balance}
\begin{document}

\title{CFAR Detection on Non-Coherently Integrated Angular Maps for mmWave Radio SLAM}

\author{
    \IEEEauthorblockN{Paolo Tosi\IEEEauthorrefmark{1}\IEEEauthorrefmark{2}, 
    Ossi Kaltiokallio\IEEEauthorrefmark{3}
    Marcus Henninger\IEEEauthorrefmark{1},
    Silvio Mandelli\IEEEauthorrefmark{1},
    and Mikko Valkama\IEEEauthorrefmark{3}
    }
    \IEEEauthorblockA{%
        \IEEEauthorrefmark{1}Nokia Bell Labs, Stuttgart, Germany,\ 
        \IEEEauthorrefmark{2}Karlsruhe Institute of Technology (KIT)\\
        \IEEEauthorrefmark{3}Tampere University, Tampere, Finland}
}%

\maketitle

\begin{abstract}
Reliable multipath detection is essential for millimeter-wave radio \gls{slam}, where propagation paths reflecting off the environment provide geometric constraints beyond the dominant line-of-sight component. 
This paper develops \gls{cfar}-based detectors for beam-swept bistatic \gls{isac}. 
Starting from conventional \gls{cfar} on the I/Q datacube, we derive a modified detector for non-coherently integrated angular power maps, where the post-integration noise statistics differ from those assumed by standard cell-averaging \gls{cfar}. 
To improve practicality, we introduce a multi-cell cell-under-test formulation and a plus-shaped training window that reduces computational cost while accounting for sidelobe structure. 
The proposed methods are evaluated on experimental 60 GHz indoor measurements using a snapshot-\gls{slam} estimator. 
Compared with a \acrfull{svd}-based detector, the proposed \gls{cfar} approaches reduce the overall positioning RMSE from 0.30 m to 0.24-0.26 m, with the largest improvement in non-line-of-sight snapshots. 
The multi-cell non-coherent detector further lowers the computational complexity without degrading positioning accuracy, demonstrating its suitability for practical multipath radio-\gls{slam}.

\glsresetall
\end{abstract}

\if\arxiv1
\blfootnote{This work has been submitted to the IEEE for possible publication. Copyright may be transferred without notice, after which this version may no longer be accessible.}
\else
\vspace{0.2cm}
\fi

\section{Introduction}
Future \gls{6g} networks are expected to support \gls{isac}, extending connectivity and positioning services towards radio-based awareness of the surrounding environment~\cite{wang2022isac_survey,3gpp_22837}. 
In the use cases studied by the \gls{3gpp}, the same transmissions are used to sense passive objects in the environment, which improves the situational awareness of the \glspl{ue}~\cite{3gpp_22837}.

Bistatic cellular \gls{slam} is one realization of this idea, in which reference signals transmitted by \glspl{bs} at known locations are used to jointly estimate the state of a \gls{ue} and the locations of environmental scattering points, commonly referred to as landmarks~\cite{shahmansoori_position_2018}. 
A radio-\gls{slam} processing chain generally consists of two stages. 
First, received measurements are converted into path estimates of \gls{aoa}, \gls{aod}, and \gls{toa}.
These estimates are then processed by a \gls{slam} algorithm to recover the \gls{ue} state and landmark positions. 
Reliable channel parameter estimation is therefore essential to the whole system. 
In particular, multipath-radio \gls{slam} exploits the \gls{nlos} components to provide additional geometric constraints, enabling positioning when the direct \gls{los} path is blocked~\cite{mendrzik2018harnessing}. 
Channel estimators must detect not only the dominant path, but also weaker multipath components with sufficiently accurate angular and delay parameters.

Many channel parameter estimators have been reported in the literature~\cite{salmi_detection_2009,alkhateeb_channel_2014,qin_angle--arrival_2019}.
They often require knowledge of the steering vectors, beamforming weights, or antenna beampatterns. 
In practical beam-sweeping systems, however, only the beam indices and their nominal pointing directions may be available, limiting the applicable channel parameter estimation methods. 
Detection directly on the angular power map is an alternative that does not require explicit knowledge of the antenna responses.
The \gls{brsrp}-based method in~\cite{rastorgueva2024millimeter} addresses this limitation by operating directly on the measured \gls{aoa}/\gls{aod} power map. 
The \mbox{rank-1} \gls{svd}-based approach provides robustness against sidelobes and noise. However, the low-rank extraction mainly retains only dominant components, whereas multipath-radio \gls{slam} benefits from retaining weaker geometrically informative returns.

Cell-averaging \gls{cfar} is well suited to detect paths on the \gls{aoa}/\gls{aod} power map.
Its conventional threshold assumes exponentially distributed single-look samples, so it can be applied to each \gls{prs} \gls{re} before integration, at the cost of one detection test per \gls{re}.
Non-coherent integration avoids this cost, but the resulting power-map cells are approximately Gaussian and the conventional threshold no longer holds.
To overcome this limitation, this paper develops a \gls{cfar} test for multipath detection in beam-swept angular maps that accounts for the post-integration Gaussian statistics.
Because a path return occupies multiple neighbouring beam pairs, the test region is extended to a multi-cell block, so the statistic is evaluated once per block rather than once per bin.
A detection is stored at the strongest cell of the \gls{cut} block, retaining the angular resolution of the single-cell test.
In addition, a plus-shaped \gls{cfar} window is proposed, to account for the directionally structured sidelobe energy along the \gls{aoa}/\gls{aod} axes. 
The proposed method is evaluated on experimental \gls{prs} measurements from the Campus Arena at Tampere University and compared with the \gls{svd}-based \gls{brsrp} estimator of~\cite{rastorgueva2024millimeter}, using the same downstream snapshot-\gls{slam} backend of~\cite{kaltiokallio_robust_2025}.
Multi-\gls{cut} non-coherent \gls{cfar} lowers positioning \gls{rmse} from $0.30$\,m to $0.25$\,m overall and from $0.38$\,m to $0.29$\,m on \gls{nlos} snapshots, while requiring about three orders of magnitude fewer detector statistic evaluations than per-\gls{re} \gls{cfar} with binary integration.

\section{System Model}
\begin{figure}[t!]
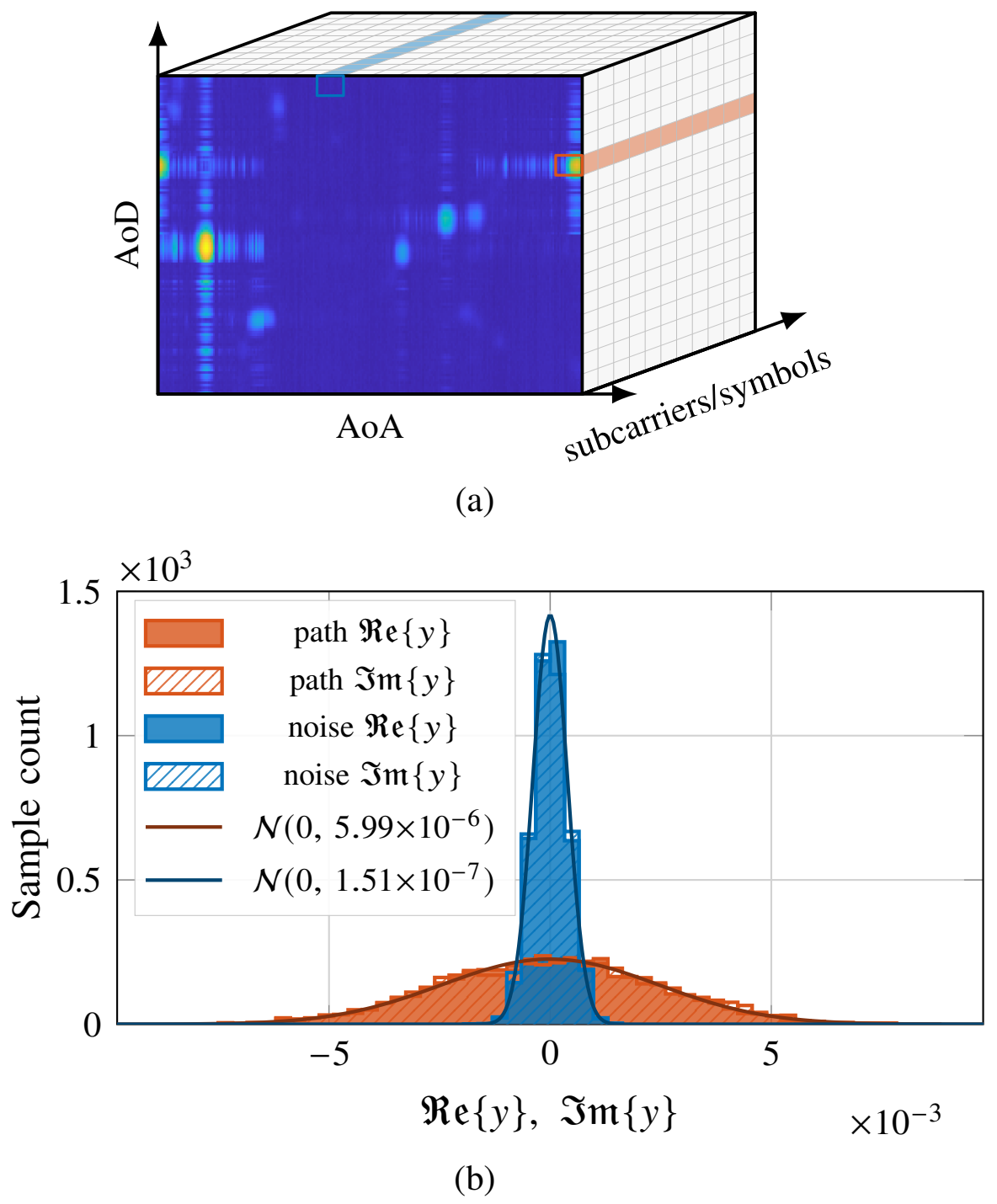

    \centering
    \providecommand{\CombinedPanelHeight}{3.6cm}%
\providecommand{\CombinedDensityWidth}{0.9\columnwidth}%

\centering

\subcaptionbox{\label{fig:datacube}}{%
    \resizebox{!}{\CombinedPanelHeight}{%
        \input{images/chan_est/datacube_3d.tikz}%
    }%
}%
\vspace{0.1em}
\subcaptionbox{\label{fig:iq-density}}{%
    \def\DensityFigWidth{0.78\columnwidth}%
    \def\DensityFigHeight{3.4cm}%
    \input{images/chan_est/angular_map_density.tikz}%
}%
    \caption{
    a) PRS I/Q datacube over AoD, AoA, and REs. The two coloured cells are one path-dominated beam pair and one noise-only beam pair.
    (b) Empirical I/Q histograms for path-dominated (orange) and noise-only (blue) beam pairs with fitted Gaussian densities.}
    \label{fig:datacube-and-density}
\end{figure}

We consider a bistatic mmWave link with a fixed \gls{bs} and a mobile UE.
For this work we focus only on the 2D azimuth plane.
\Gls{tx} and \gls{rx} hardware consists in \glspl{ula} with $\lambda/2$ spacing, where~$\lambda$ is the signal's wavelength.
For each of the recorded \gls{ue} positions in the dataset, the BS transmits a 5G NR \gls{prs} \gls{ofdm} waveform; 
\gls{prs} sequences are mapped to the the \gls{ofdm} \glspl{re} following the \gls{5g} NR specifications defined in \cite{3gpp_38211};
the \gls{ue} receives beam-swept I/Q snapshots over a discrete \gls{tx}-\gls{rx} angle grid. 
For each UE position and beam pair, received samples are organized as a three-dimensional tensor with transmit beams $N_\mathrm{TX}$, receive beams $N_\mathrm{RX}$, and fast-time samples $N_s$ as shown in \cref{fig:datacube}.
The signal propagates in a multipath environment with $N$ propagation paths.
For the $i^{\text{th}}$ TX beam and $j^{\text{th}}$ RX beam, the received signal at the $k^{\text{th}}$ subcarrier and $m^{\text{th}}$ \gls{ofdm} symbol writes as
\begin{align}
    y_{k,m}^{i,j} = \mathbf{w}_{\text{RX},j}^H \left( \mathbf{H}_{k,m} \mathbf{w}_{\text{TX},i}^*
                                                    x_{k,m}^{i,j} + \mathbf{n}_{k,m}^{i,j}
                                                    \right),
\end{align}
where $\mathbf{w}_{\text{TX},i} \in \mathbb{C}^{N_{\text{TX}}}$ and $\mathbf{w}_{\text{RX},j} \in \mathbb{C}^{N_{\text{RX}}}$ are the TX and RX beamformers, respectively, 
$x_{k,m} \in \mathbb{C}$ is the transmitted \gls{prs} sample with $|x_{k,m}| = 1 \forall k,m$, and $\mathbf{n}_{k,m}^{i,j} \in \mathbb{C}^{N_{\text{RX}}}$ is the \gls{awgn} at each RX antenna element.
The transmitted signal propagates through the multipath environment, whose effect is encoded in the effective spatial multipath channel matrix $\mathbf{H}_{k,m} \in \mathbb{C}^{N_{\text{TX}}\times N_{\text{RX}}}$ defined as \cite[eq. (5)]{Heat2016_channel_mat}
\begin{equation}
\begin{split}
    \mathbf{H}_{k,m} = \sum_{n=1}^{N} 
                            \xi_n 
                            e^{-j2\pi k\Delta f \tau_{f,n}} 
                            e^{-j2\pi mT_{\text{sym}} f_{D,n}}
                            \\
                            \times \, \mathbf{a}_{\text{RX}}(\psi_{\text{RX},n})
                            \mathbf{a}_{\text{TX}}(\psi_{\text{TX},n})^\intercal, 
\end{split}
\end{equation}
where $\Delta f$ is the subcarrier spacing and $T_{\text{sym}}$ is the \gls{ofdm} symbol duration. 
The complex path coefficient including propagation loss is denoted $\xi_n$, while $\tau_{f,n}$ and $f_{\text{D},n}$ are the path propagation delay with respect to the beginning of the received \gls{ofdm} symbol, and the Doppler shift associated with the~$n^{\text{th}}$ propagation path, respectively.
Furthermore, $\mathbf{a}_{\text{TX}}(\psi_{\text{TX},n}) \in \mathbb{C}^{N_{TX}}$ and $\mathbf{a}_{\text{RX}}(\psi_{\text{RX},n}) \in \mathbb{C}^{N_{RX}}$ are the TX and RX steering vectors, respectively.
The \Gls{prs} occupies only a subset $\mathcal{M}_{\text{RS}}$ of the \glspl{re} in the \gls{ofdm} resource grid, with cardinality $N_{\text{RS}} = \lvert\mathcal{M}_{\text{RS}}\rvert$, where each index pair $(k,m)$ identifies subcarrier~$k$ and \gls{ofdm} symbol~$m$, respectively.

\subsection{Received signal statistics}
After \gls{ofdm} demodulation, the  \glspl{re} carrying the \gls{prs} are matched-filtered against the known transmitted sequence, obtaining the complex beamformed sample $y_{k,m}^{i,j}$, whose noise-only distribution is used to derive the decision threshold and the detection statistic of the \gls{cfar} detectors of \cref{sec:detectors}.

\subsubsection{Raw IQ samples}
The square law detector sample writes as
\begin{align}
    v_{k,m}^{i,j} = \bigl\lvert y_{k,m}^{i,j} \bigr\rvert^2
                  = \mathfrak{Re}\{y_{k,m}^{i,j}\}^2 + \mathfrak{Im}\{y_{k,m}^{i,j}\}^2 .
    \label{eq:square-law}
\end{align}
When only \gls{awgn} is present, the in-phase and quadrature components of $y_{k,m}^{i,j}$ are independent zero-mean Gaussian random variables of equal variance
\begin{align}
    \mathfrak{Re}\{y_{k,m}^{i,j}\} \sim \mathcal{N}(0, \sigma^2/2), \;\;
    \mathfrak{Im}\{y_{k,m}^{i,j}\} \sim \mathcal{N}(0, \sigma^2/2),
\end{align}
so that a single square-law sample is exponentially distributed with PDF
\begin{align}
    p(v) = \frac{1}{\sigma^2} e^{-v/\sigma^2}
    \label{eq:exp-pdf}
\end{align}
and mean $\E[v]=\sigma^2$, \textit{i.e.,} the local noise power.
\Cref{fig:iq-density} compares empirical histograms of $\mathfrak{Re}\{y_{k,m}^{i,j}\}$ and $\mathfrak{Im}\{y_{k,m}^{i,j}\}$ at a noise-only and a path-dominated beam pair of the datacube. 
In the former, both components are approximated by independent $\mathcal{N}(0,\sigma^2/2)$ densities, consistent with the \gls{awgn} model.

\subsubsection{Received BRSRP power}
For each \gls{tx}-\gls{rx} beam pair $(i,j)$, the non-coherent power map $\beta \in \mathbb{R}^{N_\text{TX} \times N_\text{RX} \times 1} $ is formed by averaging over the \glspl{re} as
\begin{equation}
    \beta_{i,j} = \frac{1}{N_{\text{RS}}}
                \sum_{(k,m) \in \mathcal{M}_{\text{RS}}} v_{k,m}^{i,j},
  \label{eq:powermap}
\end{equation}
where $v_{k,m}^{i,j}$ is the square law sample of~\eqref{eq:square-law} at \gls{aod} index $i$, \gls{aoa} index $j$, subcarrier~$k$ and symbol $m$.
Cells in $\beta$ average $N_{\text{RS}}$ squared magnitudes, so the per-pixel noise law is no longer exponential.
The PDF of the sum of $V$ exponentially distributed random variables $z = \sum_{i=1}^V v_i$ is given by the Erlang distribution with shape $V$ and scale $\sigma^2$ \cite{Richards_2014},
\begin{align}
    p(z) = \frac{z^{V-1}}{(\sigma^2)^V\,(V-1)!}
         \exp\!\left(-\frac{z}{\sigma^2}\right).
    \label{eq:erlang}
\end{align}
For large $V$, the Erlang distribution approaches a Gaussian distribution with mean $V\sigma^2$ and variance $V\sigma^4$,
\begin{align} 
    z \sim \mathcal{N}\!\left(V\sigma^2,\, V\sigma^4\right).
    \label{eq:clt}
\end{align}

\section{Channel parameter estimation methods}
\label{sec:chann_est}
The technical problem consists in estimating path angles and delays from the received \gls{prs} I/Q datacube at the \gls{ue}.
In general, the problem can be separated in two steps: first, detecting the presence of a path in the \gls{aoa}/\gls{aod} plane, and then estimating its delay.
This paper focuses on the first step.
\cref{sec:cfar-background} first introduces conventional \gls{cfar} for exponentially distributed single-look power.
\cref{sec:persample} applies this test to each occupied \gls{re} and integrates the resulting binary decisions, whereas \cref{sec:proposed-cfar} non-coherently integrates the datacube and thresholds it once with a modified Gaussian statistic.

\subsection{CFAR background}
\label{sec:cfar-background}

\begin{figure}[t]
    \centering
    \begin{tikzpicture}
    [x=0.28cm,
    y=0.28cm,
    font=\small,
    axis label/.style={text height=6.2pt, text depth=0pt, inner sep=0pt},
    panel name/.style={text height=6.2pt, text depth=0pt, inner sep=0pt},
    swatch/.style={draw=black, very thin, minimum width=0.25cm,
                   minimum height=0.25cm, inner sep=0pt, outer sep=0pt,
                   anchor=west},
    legend text/.style={anchor=west, inner sep=0pt, outer sep=0pt},
    ]

    \def\N{7}%
    \def\G{3}%
    \def\T{1}%
    \def\Train{1}%

    \def\Ntest{2}%
    \def\Nguard{1}%
    \def\Ntrain{2}%
    \def\Pad{1}%

    \pgfmathsetmacro{\Ngrid}{\Ntest+2*\Nguard+2*\Ntrain+2*\Pad}

    \pgfmathsetmacro{\legendX}{\N+1.6}
    \pgfmathsetmacro{\plusX}{\N+11.0}

    \begin{scope}
        \pgfmathsetmacro{\gl}{(\N-\G)/2} \pgfmathsetmacro{\gu}{\gl+\G}
        \pgfmathsetmacro{\tl}{(\N-\T)/2} \pgfmathsetmacro{\tu}{\tl+\T}
        \pgfmathsetmacro{\trl}{\gl-\Train} \pgfmathsetmacro{\tru}{\gu+\Train}

        \fill[pattern=north east lines, pattern color=QC1] (\trl,\trl) rectangle (\tru,\tru);
        \fill[white] (\gl,\gl) rectangle (\gu,\gu);
        \fill[pattern=north west lines, pattern color=QC4] (\gl,\gl) rectangle (\gu,\gu);
        \fill[red!55] (\tl,\tl) rectangle (\tu,\tu);

        \draw[step=1, gray!55, very thin] (0,0) grid (\N,\N);
        \draw[thick] (0,0) rectangle (\N,\N);

        \node[axis label, below=6pt] at (\N/2,0) {\gls{aoa} bins};
        \node[axis label, rotate=90, anchor=south] at ([xshift=-6pt]0,\N/2) {\gls{aod} bins};
        \node[panel name, below=20pt] at (\N/2,0) {square window};
    \end{scope}

    \begin{scope}[shift={(\plusX,0)}]
        \pgfmathsetmacro{\cLo}{\Pad+\Nguard+\Ntrain} \pgfmathsetmacro{\cHi}{\cLo+\Ntest}
        \pgfmathsetmacro{\gLo}{\cLo-\Nguard}         \pgfmathsetmacro{\gHi}{\cHi+\Nguard}
        \pgfmathsetmacro{\rLo}{\gLo-\Ntrain}         \pgfmathsetmacro{\rHi}{\gHi+\Ntrain}

        \fill[pattern=north east lines, pattern color=QC1] (\rLo,\cLo) rectangle (\gLo,\cHi);
        \fill[pattern=north east lines, pattern color=QC1] (\gHi,\cLo) rectangle (\rHi,\cHi);

        \fill[pattern=north east lines, pattern color=QC3] (\cLo,\rLo) rectangle (\cHi,\gLo);
        \fill[pattern=north east lines, pattern color=QC3] (\cLo,\gHi) rectangle (\cHi,\rHi);

        \fill[pattern=north west lines, pattern color=QC4] (\gLo,\gLo) rectangle (\gHi,\gHi);
        \fill[red!55] (\cLo,\cLo) rectangle (\cHi,\cHi);

        \draw[step=1, gray!55, very thin] (0,0) grid (\Ngrid,\Ngrid);
        \draw[thick] (0,0) rectangle (\Ngrid,\Ngrid);
        
        \node[axis label, rotate=90, anchor=south] at ([xshift=-6pt]0,\Ngrid/2) {\gls{aod} bins};
        \node[axis label, below=6pt] at (\Ngrid/2,0) {\gls{aoa} bins};
        \node[panel name, below=20pt] at (\Ngrid/2,0) {plus window};
    \end{scope}

    \pgfmathsetmacro{\legStep}{2.3}
    \pgfmathsetmacro{\legTop}{\Ngrid/2+1.5*\legStep}
    \pgfmathsetmacro{\legTextX}{\legendX+1.7}

    \node[swatch, fill=red!55] at (\legendX,\legTop) {};
    \node[legend text] at (\legTextX,\legTop) {\gls{cut} $\mathcal{A}_\mathrm{test}$};

    \node[swatch, pattern=north west lines, pattern color=QC4]
        at (\legendX,{\legTop-\legStep}) {};
    \node[legend text] at (\legTextX,{\legTop-\legStep}) {guard cells};

    \node[swatch, pattern=north east lines, pattern color=QC1]
        at (\legendX,{\legTop-2*\legStep}) {};
    \node[legend text] at (\legTextX,{\legTop-2*\legStep})
        {$\mathcal{A}_\mathrm{train,1}$};

    \node[swatch, pattern=north east lines, pattern color=QC3]
        at (\legendX,{\legTop-3*\legStep}) {};
    \node[legend text] at (\legTextX,{\legTop-3*\legStep})
        {$\mathcal{A}_{\mathrm{train},2}$};
\end{tikzpicture}%
    \caption{2D \gls{cfar} windows on an \gls{aoa}/\gls{aod} bin grid. Left: square window centred on the \gls{cut} (solid), with training region $\mathcal{A}_\mathrm{train}$ (striped). Right: plus-shaped window with an $N_\mathrm{test}=2$ \gls{cut} block and orthogonal training strips $\mathcal{A}_{\mathrm{train},1}$ (\gls{aod}-aligned) and $\mathcal{A}_{\mathrm{train},2}$ (\gls{aoa}-aligned) providing separate noise references.}
    \label{fig:cfar-windows}
\end{figure}
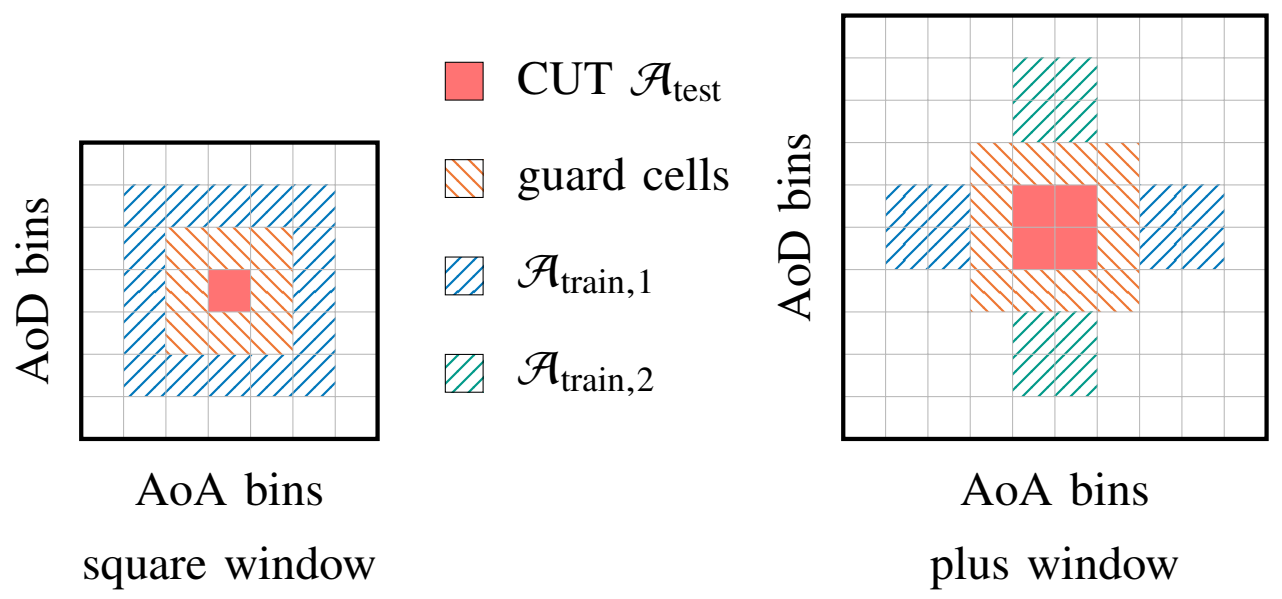

To determine which bin of a radar image is relevant, \textit{i.e.,} corresponds to a peak, a \acrfull{cfar} detector is commonly employed \cite{Richards_2014}.
For each \acrfull{cut}, the detector compares power in a test region $\mathcal{A}_\mathrm{test}$ to a data-dependent threshold set from a local interference estimate.
In the standard \gls{cfar} geometry, $\mathcal{A}_\mathrm{test}$ is surrounded by a guard band and a training region $\mathcal{A}_\mathrm{train}$ whose cells supply the reference statistic. 
The guard band prevents energy correlated to the \gls{cut} from biasing the noise estimate.
\cref{fig:cfar-windows}~(left) illustrates an example 2D square \gls{ca-cfar-short} window.
For each \gls{cut}, detection is cast as a binary hypothesis test
\begin{align}
    \mathcal{H}_0 &: \text{the cell contains noise only}, \notag \\
    \mathcal{H}_1 &: \text{the cell contains a path return plus noise},
    \label{eq:hypotheses}
\end{align}
where the \gls{cut} statistic is compared against a threshold chosen to maintain a constant probability of false alarm $P_\mathrm{fa} = P\!\left(\text{decide } \mathcal{H}_1 \mid \mathcal{H}_0\right)$.
\gls{cfar} detectors operate on the output of a square law detector, which under $\mathcal{H}_0$ is exponentially distributed when the received in-phase and quadrature components are zero-mean Gaussian \cite{Richards_2014}.

Denoting by $R = |\mathcal{A}_\mathrm{train}|$ the number of training cells and by $v_\mathrm{CUT}$ the square-law power~\eqref{eq:square-law} at the \gls{cut}, \gls{cfar} detectors compute the noise reference as the cell average
    $\hat{\mu}_\mathrm{train} = \frac{1}{R}\sum_{r\in\mathcal{A}_\mathrm{train}} v_r,$
and declares a detection when
\begin{align}
    v_\mathrm{CUT}
      \;\underset{\mathcal{H}_0}{\overset{\mathcal{H}_1}{\gtrless}}\;
      \eta_\mathrm{CFAR},
    \quad
    \eta_\mathrm{CFAR} = \alpha\,\hat{\mu}_\mathrm{train},
    \quad
    \alpha = R\!\left(P_\mathrm{fa}^{-1/R} - 1\right),
\label{eq:cfar-threshold}
\end{align}
where $\alpha$ is fixed from the exponential model~\eqref{eq:exp-pdf} under $\mathcal{H}_0$ so that $P_\mathrm{fa} = P\!\left(v_\mathrm{CUT} > \eta_\mathrm{CFAR} \mid \mathcal{H}_0\right)$~\cite{Richards_2014}.

\subsection{Proposed CFAR detectors}
\label{sec:detectors}

\subsubsection{Per-resource-element CFAR with binary integration}
\label{sec:persample}

Classical \gls{cfar} can be applied directly on each frequency--time slice of the datacube: for each beam pair $(i,j)$ and each $(k,m)\in\mathcal{M}_{\text{RS}}$, demodulation yields one square-law sample  at a fixed \gls{aoa}/\gls{aod} cell. As per~\eqref{eq:exp-pdf}, this single-look statistic is exponentially distributed.
The \gls{ca-cfar-short} rule~\eqref{eq:cfar-threshold} is therefore evaluated once per \gls{re}, before any integration across $\mathcal{M}_{\text{RS}}$.
The observed sample distributions of a noise-only cell and a path-dominated cell on measured data is shown in \cref{fig:iq-density}.

A single \gls{re} carries only a fraction of the signal's energy, so the per-slice decisions are not used individually but accumulated over the datacube.
Applying~\eqref{eq:cfar-threshold} to each \gls{re} of a beam pair $(i,j)$ yields the binary decision
\begin{equation}
  s_{k,m}^{i,j} = \mathbf{1}\!\left( v_{k,m}^{i,j} > \eta_{\mathrm{CFAR},k,m}^{i,j} \right),
  \label{eq:slice-decision}
\end{equation}
which are combined by binary integration into the vote count
\begin{equation}
  S_{i,j} = \sum_{(k,m)\in\mathcal{M}_{\text{RS}}} s_{k,m}^{i,j}.
  \label{eq:vote-count}
\end{equation}
Let us define $q$ as the false-alarm probability on each slice.
Under $\mathcal{H}_0$, each of the $N_{\text{RS}}$ decisions in~\eqref{eq:vote-count} is a Bernoulli trial of success probability $q$, by construction of $\eta_{\text{CFAR}}$ in~\eqref{eq:cfar-threshold}.
Treating the decisions as independent across \glspl{re}, $S_{i,j}$ is binomial with $N_{\text{RS}}$ trials, mean $N_{\text{RS}}q$ and variance $N_{\text{RS}}q(1-q)$, and for $N_{\text{RS}} \gg 1$ the central limit theorem applies.
The standardised count
\begin{equation}
  \tilde{t}_{i,j} = \frac{S_{i,j}-N_{\text{RS}}\,q}{\sqrt{N_{\text{RS}}\,q\,(1-q)}}
  \;\underset{\mathcal{H}_0}{\overset{\mathcal{H}_1}{\gtrless}}\;
  Q^{-1}\!\left(P_\mathrm{fa}\right)
  \label{eq:binary-integration}
\end{equation}
is therefore approximately standard normal under $\mathcal{H}_0$, for any $q$.
A path return raises the detection probability of the individual slices above $q$, therefore increasing the mean of $S_{i,j}$ above $N_{\text{RS}}q$. Then $\mathcal{H}_1$ corresponds to large $\tilde{t}_{i,j}$.
Its false-alarm probability is $P\!\left(\tilde{t}_{i,j} > \gamma \mid \mathcal{H}_0\right) = Q(\gamma)$, so the Gaussian quantile $\mbox{$\gamma = Q^{-1}(P_\mathrm{fa})$}$ in~\eqref{eq:binary-integration} controls the rate $P_\mathrm{fa}$ for the datacube.
The detector operates with two false-alarm rates: $q$ on each slice, which sets how many votes noise is expected to produce, and $P_\mathrm{fa}$ on the datacube, which sets how many votes are needed to declare a path.

The two rates have to be chosen jointly, and the binomial model behind~\eqref{eq:binary-integration} assumes decisions that are independent across \glspl{re}, whereas neighboring ones observe the same channel realization. 
The statistic and threshold test of~\eqref{eq:cfar-threshold} are evaluated once per \gls{re} and beam pair, which amounts to a complete scan for each of the $N_{\text{RS}}$ \gls{aoa}/\gls{aod} slices. 

\subsubsection{Non-coherent integration in the \gls{aoa}/\gls{aod} plane}
\label{sec:proposed-cfar}

As a more efficient alternative, we integrate the power over the \glspl{re} first and threshold once.
The non-coherent power map $\beta$ of~\eqref{eq:powermap} averages  square-law samples of each beam pair, so detection reduces to a single \gls{cfar} scan of the \gls{aoa}/\gls{aod} plane.
Integrating the power changes the noise law.
By~\eqref{eq:erlang} and \eqref{eq:clt} the cells of $\beta$ are approximately Gaussian, so the threshold in~\eqref{eq:cfar-threshold} does not achieve the nominal false-alarm rate.
This motivates the modified Gaussian $Q$-function test derived next.

The test and training reference levels are now formed from the sample moments of the corresponding regions of the \gls{cfar} window on $\beta$.
For a region $\mathcal{A}_s$, $s\in\{\mathrm{test},\mathrm{train}\}$, of cardinality $\lvert \mathcal{A}_s\rvert$; the sample mean and variance are
\begin{equation}
\begin{aligned}
  \hat{\mu}_s &= \frac{1}{\lvert \mathcal{A}_s\rvert}
                \sum_{(i,j)\in\mathcal{A}_s} \beta_{i,j},
  \\
  \hat{\sigma}^2_s &= \frac{1}{\lvert \mathcal{A}_s\rvert}
                \sum_{(i,j)\in\mathcal{A}_s} \left(\beta_{i,j}-\hat{\mu}_s\right)^2 .
\end{aligned}
  \label{eq:region-moments}
\end{equation}
Under $\mathcal{H}_0$ cells in both regions sample the noise-only population in~\eqref{eq:clt}, so both estimates are approximately Gaussian and centred on the common mean $\sigma_s^2$.
Under $\mathcal{H}_1$ the test region contains path energy, which increases $\hat{\mu}_\mathrm{test}$ with respect to the training mean.
Detection is therefore declared on the difference between the test and training means, normalized by the training standard deviation of training cells
\begin{equation}
  \frac{\hat{\mu}_\mathrm{test}-\hat{\mu}_{\mathrm{train}}}
           {\hat{\sigma}_{\mathrm{train}}}
      \;\underset{\mathcal{H}_0}{\overset{\mathcal{H}_1}{\gtrless}}\;
      Q^{-1}\!\left(P_\mathrm{fa}\right),
  \label{eq:cfar-statistic}
\end{equation}
where $Q(x) = \frac{1}{\sqrt{2\pi}}\int_x^{\infty} e^{-u^2/2}\,\mathrm{d}u$ is the Gaussian tail probability.
Since $\hat{\mu}_\mathrm{test}$ and $\hat{\mu}_\mathrm{train}$ are averages of the approximately Gaussian cells in~\eqref{eq:clt}, the standardised statistic is itself approximately standard normal under $\mathcal{H}_0$; the Gaussian quantile $Q^{-1}(P_\mathrm{fa})$ therefore replaces the inverse-exponential threshold of standard \gls{cfar} in~\eqref{eq:cfar-threshold}.

\subsection{CFAR window design}
The generic detectors of~\eqref{eq:cfar-threshold} and \eqref{eq:cfar-statistic} leave open the design choices of the shape of the test and of the training regions.

\subsubsection{Test region} A single path return is, in general, sampled by several cells.
Evaluating every single-bin \gls{cut} in the angular map is therefore complex and possibly redundant, as a single path spanning several cells yields several detections.
The primary motivation for our multi-cell test region is thus computational complexity. We group $N_\mathrm{test}\times N_\mathrm{test}$ cells into one test block and slide it across the angular map.
The local statistic and threshold are evaluated once per block rather than at every bin, reducing the number of operations by a factor set by the block stride, up to the block area $N_\mathrm{test}^2$ for non-overlapping placement.
Furthermore, averaging over the test region $\hat{\mu}_\mathrm{test}$ makes the detector less sensitive to false alarms from per-cell fluctuations due to noise and sidelobes.
When a block passes the detection test, the detection is assigned to the strongest pixel within the group, so the original per-bin resolution of the map is retained.

\subsubsection{Training region} In general, the shape of the \gls{cfar} window, together with the number of guard and training cells, is a design choice matched to the characteristics of the returns in the radar image.
On the angular map, sidelobes from strong returns extending along the \gls{aod} and \gls{aoa} axes are significantly stronger than the noise and may mask weak returns.
We choose to align the training region with sidelobes, using a plus-shaped window built from two orthogonal strips, an \gls{aod}-aligned vertical band $\mathcal{A}_{\mathrm{train},1}$ and an \gls{aoa}-aligned horizontal band $\mathcal{A}_{\mathrm{train},2}$, as shown in \cref{fig:cfar-windows}~(right).
Each strip yields its own moments $\hat{\mu}_{\mathrm{train},k}$ and $\hat{\sigma}_{\mathrm{train},k}$, $k\in\{1,2\}$, from~\eqref{eq:region-moments}, 
and the test statistic~\eqref{eq:cfar-statistic} is taken by comparing the cut against the training strip with less power, \textit{i.e.,} comparing the threshold using the maximum of the two normalized contrasts
\begin{equation}
  \max_{k\in\{1,2\}}
      \frac{\hat{\mu}_\mathrm{test}-\hat{\mu}_{\mathrm{train},k}}
           {\hat{\sigma}_{\mathrm{train},k}}
      \;\underset{\mathcal{H}_0}{\overset{\mathcal{H}_1}{\gtrless}}\;
      Q^{-1}\!\left(P_\mathrm{fa}\right).
  \label{eq:block-cfar-plus}
\end{equation}

\subsection{Raw detection clustering}
\label{sec:peak-clustering}
A single path return typically generates several neighboring cells above threshold.
The detection map is therefore reduced to one candidate per path before detections are converted to angle and delay estimates.
Let $\mathcal{D}$ denote the set of cells declared by the \gls{cfar} detector.
Detections in $\mathcal{D}$ are clustered by single linkage: two cells $(i,j)$ and $(i',j')$ are merged when their index distance satisfies $\lVert(i,j)-(i',j')\rVert_2 \leq N_\mathrm{test}/\sqrt{2}$, so that a chain of adjacent alarms along a lobe forms one group.
This yields a partition $\{\mathcal{C}_n\}_{n=1}^{N_\mathrm{path}}$, where each cluster $\mathcal{C}_n\subset\mathcal{D}$ collects detections associated with one tentative path.
Each cluster is represented by its strongest cell
\begin{equation}
  (\hat{\imath}_n,\hat{\jmath}_n)
  = \arg\max_{(i,j)\in\mathcal{C}_n} \beta_{i,j}.
  \label{eq:cluster-peak}
\end{equation}

\section{Angle and Delay estimation}
\label{sec:angle-delay}
\subsection{Angle estimation}
\label{sec:refinement}

The angular path return is quantized to the discrete \gls{aoa}/\gls{aod} grid.
If the true angle does not lie exactly in the $(\hat{\imath}_n,\hat{\jmath}_n)$ grid point, then the estimation error is at most half a cell.
The path main lobe is, in general, sampled by several cells and its shape can be exploited to interpolate the peak position.
For each discrete path estimate, we fit a 2D Gaussian with a constant pedestal to the power map $\beta$ over a fixed square patch of $N_\mathrm{fit}+1$ cells per side, centred on the retained peak $(\hat{\imath}_n,\hat{\jmath}_n)$
\begin{equation}
  \bigl(\hat{D},\hat{A},\hat{\boldsymbol{\mu}}_n,\hat{\mathbf{C}}_n\bigr)
  = \arg\min \!\!\sum_{(i,j)\in\mathcal{W}_n} \!\!
    \Bigl(\beta_{i,j} - D - A\,e^{-\frac{1}{2}\mathbf{d}_{i,j}^\intercal \mathbf{C}^{-1}\mathbf{d}_{i,j}}\Bigr)^{2},
  \label{eq:gauss-fit}
\end{equation}
where $\mathbf{d}_{i,j} = [i\;\, j]^\intercal - \boldsymbol{\mu}$ is the offset of cell $(i,j)$ from the lobe centre $\boldsymbol{\mu}$, expressed in fractional cell indices, and $\mathbf{C}$ is the covariance matrix that sets width and orientation of the lobe. 
Parameters are estimated jointly by nonlinear least squares.
The pedestal $D$ accounts for the local noise floor, and $A$ is the amplitude of the Gaussian lobe.
The fitted mean $\hat{\boldsymbol{\mu}}_n$ is a fractional cell index: the refined angles $\hat{\psi}_{\mathrm{TX},n}$ and $\hat{\psi}_{\mathrm{RX},n}$ are obtained by linear interpolation between adjacent beam values.

\subsection{Delay estimation and sidelobe removal}
\label{sec:delay}
Each detected tentative path is later associated with a time-of-arrival estimate $\hat{\tau}$ obtained by correlating the received waveform against stored \gls{prs} references.
For the $n$-th propagation path at the \gls{rx}, the total physical propagation delay from the \gls{tx} is denoted by $\tau_n = \tau_c + \tau_{f,n}$, where $\tau_c$ is the coarse delay between received \gls{ofdm} symbol and transmit time.
Considering a \gls{rx} with unknown clock bias $b_{\text{UE}}$, the biased propagation delay is
\begin{align}
    \tau_n^b = \tau_n - b_{\text{UE}} = \tau_c + \tau_{f,n} - b_{\text{UE}}.
\end{align}
Delay estimation is performed in two steps.
The coarse delay $\tau_c$ is obtained from the time-domain I/Q of the detected beam pair $(\hat{\imath}_n,\hat{\jmath}_n)$, as the lag that maximises the cross-correlation of the received waveform with a stored \gls{prs} replica.
The fine delay $\tau_{f,n}$ appears as the linear phase $e^{-j2\pi k\Delta f\tau}$ across subcarriers.
It is obtained by a sub-sample grid search that maximises the coherent correlation of the least-squares \gls{prs} channel estimate on the detected beam pair after compensating for the coarse delay.
Sidelobe filtering then discards weaker returns that are close in angle and delay to a stronger peak.
For more details regarding this procedure, please refer to \cite{rastorgueva2024millimeter}.

\section{Experimental Results}
\begin{table*}[t!]
    \centering
    \caption{Comparison of detector approaches: best snapshot-\gls{slam} positioning
             \gls{rmse} (distance), detector-stage statistic evaluations per
             snapshot $N_\mathrm{eval}$ and threshold time $T_\mathrm{thr}$. 
             Time $T_\mathrm{det}$ accounts for the map formation and the vote gate.
             All \gls{cfar} rows share the same nested-loop MATLAB implementation over each \gls{cut} in the AOA/AOD plane.}
    \label{tab:three-way}
    \begin{tabular}{lrrrrrr}
      \toprule
      Approach & 
      RMSE$_\mathrm{LOS}$ [m] & RMSE$_\mathrm{NLOS}$ [m] & RMSE$_\mathrm{ALL}$ [m] &
      $N_\mathrm{eval}$ & $T_\mathrm{thr}$ [s] & $T_\mathrm{det}$ [s] \\
      \midrule
      Baseline: \gls{svd} \gls{brsrp} &  0.272 & 0.378 & 0.303 & --- & --- & --- \\
      Per-\gls{re} \gls{cfar} + vote gate & 
      \textbf{0.223} & 0.294 & \textbf{0.243} & $5.03\times10^{7}$ & 67.7 & 68.1 \\
      NC \gls{cfar} (4 CUT cells, plus window) & 0.229 & \textbf{0.293} & 0.247 & $7.75\times10^{3}$ & 0.0303 & 1.371 \\
      NC \gls{cfar} (single \gls{cut}, plus window) & 0.228 & 0.313 & 0.252 & $3.18\times10^{4}$ & 0.1006 & 1.442 \\
      NC \gls{cfar} (4 CUT cells, square window) & 0.230 & 0.342 & 0.263 & $7.75\times10^{3}$ & \textbf{0.0198} & \textbf{1.349} \\
      \bottomrule
    \end{tabular}
  \end{table*}

\subsection{Experimental Setting}
\label{sec:campaign}

Experiments use the open 60\,GHz indoor dataset collected at the Hervanta campus, Tampere University~\cite{rastorgueva2024millimeter}.
The dataset consists in a set of bistatic measurements where a fixed \gls{bs} illuminates a \gls{ue} which moves along a predefined trajectory while both sides perform a mechanical beam sweep~\cite{rastorgueva2024millimeter}.
The hardware consists of Sivers EVK06002 front ends (57--71\,GHz) with \glspl{ula} with $16$ horizontal elements, coax-based time synchronisation for delay reference, and a pan-tilt unit on the \gls{rx} side to cover the full azimuth sector.
At each of the 43 trajectory points (nominal spacing $\approx 0.5$\,m), the \gls{bs} transmits a 5G NR \gls{prs} \gls{ofdm} waveform with $400$\,MHz occupied bandwidth.
The \gls{ue} records complex I/Q for every \gls{tx}-\gls{rx} beam pair in the sweep.
Thirty-two positions are predominantly \gls{los} with respect to the \gls{bs} and thirteen are \gls{nlos}, \textit{i.e.}, when the \gls{ue} moves behind columns or corners.

\begin{figure}[t]
    \centering
    \def\SlamTrajWidth{0.9\linewidth}%
    \def\SlamTrajHeight{0.60\linewidth}%
    \noindent
\begin{tikzpicture}[baseline]
\node[inner sep=0pt] {%
\begin{tabular}{@{}c@{\,}l@{\hspace{0.7em}}c@{\,}l@{\hspace{0.7em}}c@{\,}l@{\hspace{0.7em}}c@{\,}l@{}}
\begin{tikzpicture}[baseline=-0.35ex]
  \draw[blue, line width=0.9pt]
    (0,-0.11) -- (0.10,0.07) -- (-0.10,0.07) -- cycle;
\end{tikzpicture}
& \footnotesize TX
&
\begin{tikzpicture}[baseline=-0.15ex]
  \draw[blue, line width=1pt] (0,0) -- (0.30,0);
\end{tikzpicture}
& \footnotesize UE trajectory
&
\begin{tikzpicture}[baseline=-0.35ex]
  \draw[black, line width=1.0pt]
    (-0.07,-0.07) -- (0.07,0.07)
    (-0.07, 0.07) -- (0.07,-0.07);
\end{tikzpicture}
& \footnotesize est.\ pos.
&
\begin{tikzpicture}[baseline=-0.35ex]
  \draw[red, line width=0.9pt]
    (-0.08,0) -- (0.08,0)
    (0,-0.08) -- (0,0.08);
\end{tikzpicture}
& \footnotesize landmarks
\end{tabular}%
};
\end{tikzpicture}%
\\[2pt]
      \centering
      \providecommand{\SlamTrajImage}{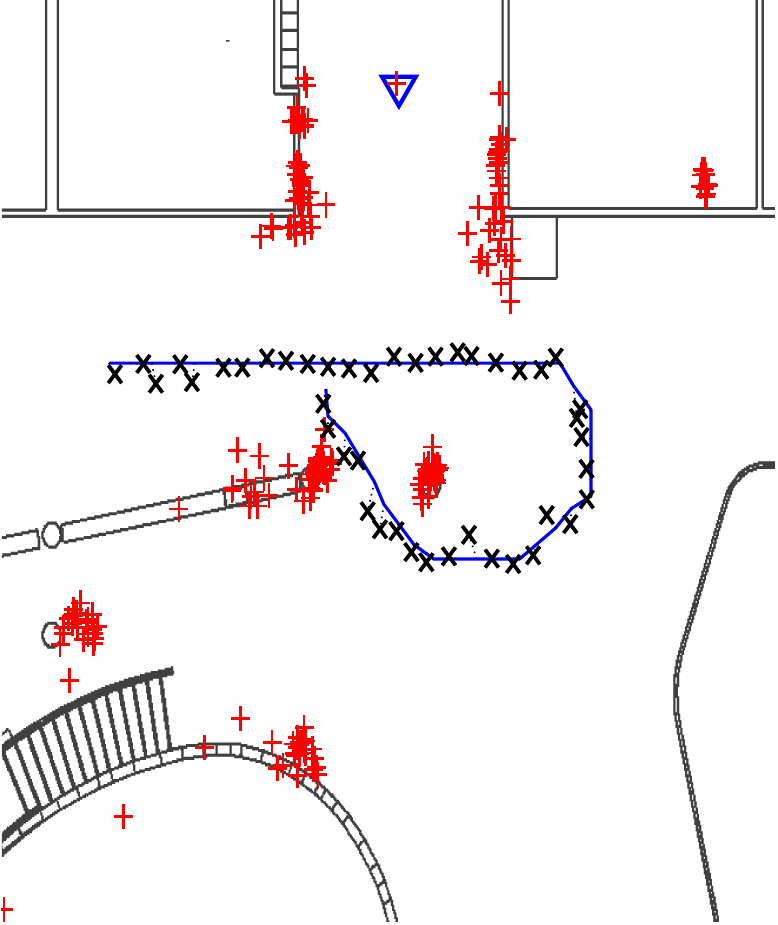}%
\input{images/results/slam_trajectory_panel.tikz}%
      \caption{Snapshot-\gls{slam} positioning outcome on the measurement floor plan for non-coherent \gls{cfar} (4 CUT cells, 'plus' window).}
    \label{fig:slam-trajectory-multi-cut}
  \end{figure}
  
\subsection{Evaluation}
\label{sec:evaluation}

All detectors are compared using the same raw I/Q waveforms and forming the same \gls{aoa}/\gls{aod}/\gls{re} datacube.
Furthermore, they share the same downstream stages: fixed-window Gaussian refinement, \gls{prs}-based delay estimation and sidelobe filtering.
At each trajectory point, one complete \gls{tx}-\gls{rx} beam sweep forms a snapshot. Each detector extracts a set of path measurements from that snapshot, comprising angles, delay and power. The regularised robust snapshot-\gls{slam} estimator of~\cite{kaltiokallio_robust_2025} processes these paths independently for each snapshot and returns one \gls{ue} position and a set of landmark positions.
The detectors are compared over the positioning \gls{rmse} in the azimuthal plane.

We test three groups of angular detectors:
\begin{itemize}
  \item Baseline \gls{svd} \gls{brsrp}, reference from~\cite{rastorgueva2024millimeter};
  \item Proposed per-\gls{re} \gls{cfar}, per-slice detector with binary integration of \cref{sec:persample};
  \item Proposed non-coherent integration detector of \cref{sec:proposed-cfar}.
\end{itemize}

The non-coherent \gls{cfar} detector is tested in three configurations: using the plus-shaped window with  single-\gls{cut}, multi-\gls{cut} and using a standard square window with the multi-\gls{cut}.
The per-\gls{re} detector uses a single-\gls{cut} and the same plus-shaped window.
For each compared approach we sweep the detector hyperparameters, \textit{i.e.}, nominal false-alarm rate $P_\mathrm{fa}$ and the per-slice rate $q$ where applicable, and retain the operating point with the lowest \gls{rmse}.

\subsection{Results}
\label{sec:results}

\subsubsection{Positioning accuracy} 

Table~\ref{tab:three-way} reports the best snapshot positioning \gls{rmse} for each approach.
The baseline \gls{svd} \gls{brsrp} approach retains the dominant returns of each angular map, and is therefore reliable against false alarms from sidelobes and noise.
It is well suited to a \gls{los}-based \gls{slam} estimator, but it is less sensitive to the weak secondary returns that carry geometric information of the multipath environment.
The \gls{cfar} detectors achieve an overall \gls{rmse} of $0.24$--$0.26\,$m, improving over the $0.30\,$m of \gls{svd} \gls{brsrp}.
The gain is larger on the \gls{nlos} snapshots, where the error falls to $0.29$-$0.31\,$m against $0.38\,$m of the \gls{svd} \gls{brsrp}.
The per-\gls{re} detector reaches similar accuracy in this evaluation ($0.243,$m against $0.247,$m over 43 snapshots).
Reducing the test region from four cells to one leaves the \gls{rmse} close to the multi-\gls{cut} detector, but requires the processing of a significant larger amount of detections.

By using a conventional square \gls{cfar} window, in the multi-\gls{cut} configuration,  the \gls{los} \gls{rmse} remains essentially unchanged ($0.23\,$m against $0.22\,$m) but raises the \gls{nlos} \gls{rmse} from $0.29\,$m to $0.34\,$m.
Beam-pattern sidelobes extend in the angular map around a strong return.
A square training region does not consider the angular return structure and sideobes may mask weak returns.
The plus-shaped window preserves the contrast with the noise reference by selecting the larger of the two per-strip statistics in~\eqref{eq:block-cfar-plus}, \textit{i.e.,} by comparing the \gls{cut} against the strip with less power.

\subsubsection{Sensitivity to the operating point}

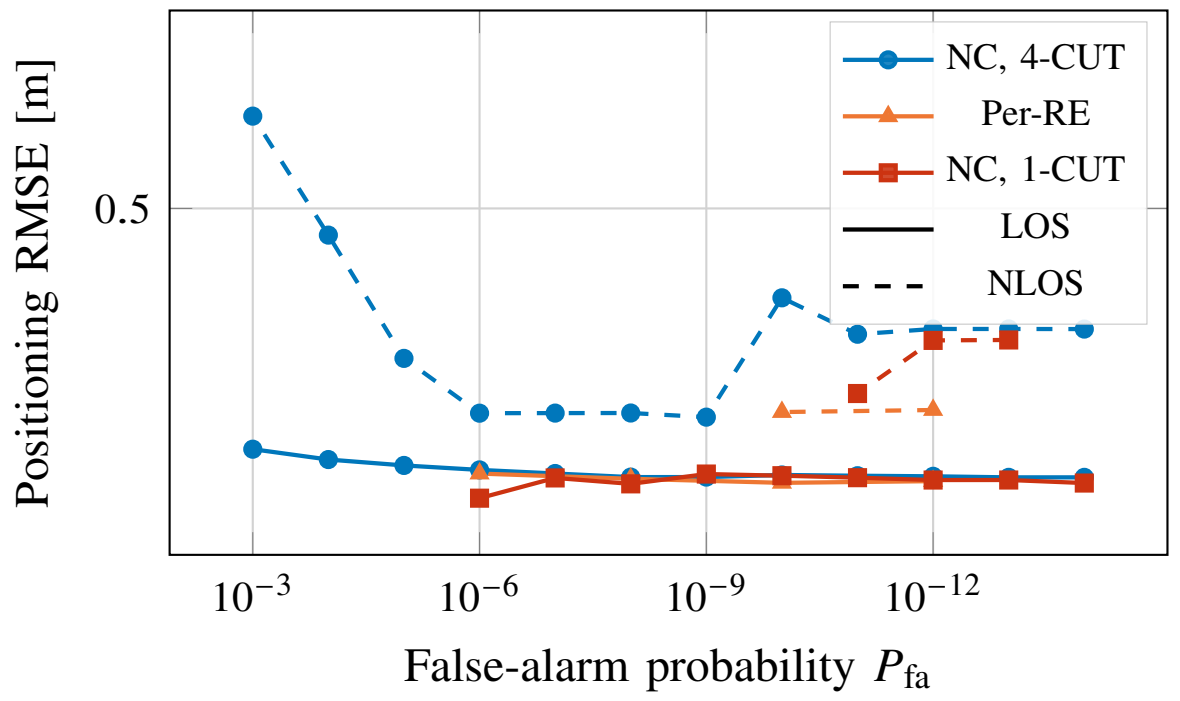
\begin{figure}[t]
    \centering
    \providecommand{\RmsePfaFigWidth}{6.6cm}%
\providecommand{\RmsePfaFigHeight}{3.6cm}%
\providecommand{\RmsePfaData}{data/rmse_vs_pfa}%
\providecommand{\RmsePfaShowBands}{0}%
\newcommand{\RmsePfaAddBand}[3]{%
    \addplot[name path=#1lo, draw=none, forget plot]
        table[x=pfa, y=lo, col sep=space] {#2};%
    \addplot[name path=#1hi, draw=none, forget plot]
        table[x=pfa, y=hi, col sep=space] {#2};%
    \addplot[fill=#3, fill opacity=0.12, draw=none, forget plot]
        fill between[of=#1lo and #1hi];%
}
\newcommand{\RmsePfaPlotDet}[4]{%
    \addplot[#1, thick, solid, mark=#2, mark size=1.4pt, forget plot,
             mark options={solid, fill=#1, draw=#1},
             restrict y to domain=0:2]
        table[x=pfa, y=rmse, col sep=space] {#3};%
    \addplot[#1, thick, dashed, mark=#2, mark size=1.4pt, forget plot,
             mark options={solid, fill=#1, draw=#1},
             restrict y to domain=0:2]
        table[x=pfa, y=rmse, col sep=space] {#4};%
}
\begin{tikzpicture}
\begin{axis}[
    width=\RmsePfaFigWidth,
    height=\RmsePfaFigHeight,
    scale only axis,
    tick label style={font=\footnotesize},
    label style={font=\small},
    xmode=log,
    xlabel={False-alarm probability $P_\mathrm{fa}$},
    ylabel={Positioning RMSE [m]},
    x dir=reverse,
    log basis x=10,
    xtick={1e-3,1e-6,1e-9,1e-12},
    ymin=0.15,
    ymax=0.7,
    ytick={0.5,1,1.5},
    clip=true,
    grid=major,
    major grid style={gray!35, thin},
    minor y tick num=1,
    unbounded coords=jump,
    legend style={
        font=\scriptsize,
        fill=white,
        fill opacity=0.88,
        text opacity=1,
        draw=gray!45,
        line width=0.2pt,
        legend columns=1,
        inner sep=2pt,
        at={(0.98,0.98)},
        anchor=north east,
        row sep=0.4pt,
    },
]
\ifthenelse{\equal{\RmsePfaShowBands}{1}}{%
    \RmsePfaAddBand{propL}{\RmsePfaData/proposed_los.txt}{QC1}%
    \RmsePfaAddBand{propN}{\RmsePfaData/proposed_nlos.txt}{QC1}%
    \RmsePfaAddBand{scutL}{\RmsePfaData/scut_los.txt}{QC5}%
    \RmsePfaAddBand{scutN}{\RmsePfaData/scut_nlos.txt}{QC5}%
}{}
\RmsePfaPlotDet{QC1}{*}{\RmsePfaData/proposed_los.txt}{\RmsePfaData/proposed_nlos.txt}
\RmsePfaPlotDet{QC4}{triangle*}{\RmsePfaData/perre_los.txt}{\RmsePfaData/perre_nlos.txt}
\RmsePfaPlotDet{QC5}{square*}{\RmsePfaData/scut_los.txt}{\RmsePfaData/scut_nlos.txt}
\addlegendimage{sharp plot, QC1, thick, solid, mark=*, mark size=1.4pt,
                mark options={solid, fill=QC1, draw=QC1}}
\addlegendentry{NC, 4-CUT}
\addlegendimage{sharp plot, QC4, thick, solid, mark=triangle*, mark size=1.4pt,
                mark options={solid, fill=QC4, draw=QC4}}
\addlegendentry{Per-RE}
\addlegendimage{sharp plot, QC5, thick, solid, mark=square*, mark size=1.4pt,
                mark options={solid, fill=QC5, draw=QC5}}
\addlegendentry{NC, 1-CUT}
\addlegendimage{sharp plot, black, thick, solid}
\addlegendentry{LOS}
\addlegendimage{sharp plot, black, thick, dashed}
\addlegendentry{NLOS}
\end{axis}
\end{tikzpicture}%
    \caption{Snapshot positioning \gls{rmse} against the operating $P_\mathrm{fa}$ for the non-coherent detector, the per-\gls{re} detector and the single-\gls{cut} non-coherent detector, separated into \gls{los} (solid) and \gls{nlos} (dashed) snapshots. RMSE values exceeding axes bounds are not shown.}    \label{fig:rmse-vs-pfa}
\end{figure}

\cref{fig:rmse-vs-pfa} shows the \gls{rmse} as a function of the false-alarm probability $P_\mathrm{fa}$.
For the non-coherent detector, raising the threshold (\emph{i.e.,} decreasing $P_\mathrm{fa}$) lowers the \gls{nlos} \gls{rmse} from $0.59\,$m at $P_\mathrm{fa}=10^{-3}$ to a minimum of $0.29\,$m at $10^{-9}$, and on \gls{los} snapshots from $0.26\,$m to $0.23\,$m.
Below $10^{-9}$, \gls{nlos} error rises again, as weak multipath returns carrying geometric information fall below threshold, and the \gls{slam} estimator is unable to solve the multipath problem.
Between $10^{-6}$ and $10^{-9}$ both of its curves are nearly flat, so this represents the practical working band.
The per-\gls{re} and single-\gls{cut} detectors are more sensitive to tuning parameters.
The per-\gls{re} detector obtains the best performance in a narrow band around $10^{-10}$-$10^{-12}$, and the single-\gls{cut} between $10^{-11}$ and $10^{-13}$.

\subsubsection{Computational cost}

\balance

We compare the number of statistic evaluations per snapshot, $N_\mathrm{eval}$ and the threshold time $T_\mathrm{thr}$. 
Furthermore, $T_\mathrm{det}$ accounts for the map formation and, for the per-\gls{re} detector only, the vote count~\eqref{eq:vote-count}.
All detectors use the same nested-loop implementation of the threshold test over cells of the \gls{aoa}/\gls{aod} plane.
The per-\gls{re} detector applies the threshold test once per occupied \gls{prs} resource element on every beam pair, which requires $N_\mathrm{eval}=5.03\times10^{7}$, $T_\mathrm{thr}=67.7\,\mathrm{s}$ and $T_\mathrm{det}=68.1\,\mathrm{s}$.
Non-coherent integration first collapses the datacube onto a single map, so the test is evaluated only over beam pairs.
The map formation time is $T_\mathrm{map}=1.33\,\mathrm{s}$.
A single-cell \gls{cut} with unit stride evaluates the statistic at every bin, which requires $N_\mathrm{eval}=3.18\times10^{4}$ and $T_\mathrm{thr}=0.101\,\mathrm{s}$.
With a four-cell \gls{cut} the detection time improves to $N_\mathrm{eval}=7.75\times10^{3}$ and $T_\mathrm{thr}=30.3\,\mathrm{ms}$.
The square-window variant uses the same 4-cell \gls{cut}, but does not require computing the maximum of the two training strips ($N_\mathrm{eval}=7.75\times10^{3}$, $T_\mathrm{thr}=19.8\,\mathrm{ms}$).
In terms of computational complexity, non-coherent integration requires $10^3$ less statistic evaluations than per-\gls{re} detection, leading  to a considerably lower threshold time and total detector time including map formation.

\label{sec:exp-results}

\section{Conclusion}
In this paper we formulated conventional \gls{cfar} detection on the I/Q datacube for mmWave bistatic \gls{isac} and derived a modified detector for non-coherently integrated angular power maps, including a multi-\gls{cut} extension for reduced complexity.
On experimental data, the proposed detectors reduced the overall positioning \gls{rmse} from the \gls{svd} baseline by approximately $20\%$, with the largest improvement in \gls{nlos} conditions.
The proposed detector is also able to detect weak multipath returns that are fundamental for solving the \gls{slam} problem accurately.

\section{acknowledgments}
Paolo Tosi has been supported by the European Commission through the ISLANDS project (grant agreement no. 101120544).

\bibliographystyle{IEEEtran}
\bibliography{referencing/bibliography}

\end{document}